\documentclass[10pt,letterpaper,twocolumn]{article} 

\usepackage{ol2}
\usepackage[draft]{hyperref}
\usepackage{amsmath}
\usepackage{graphicx}

\begin{document}

\twocolumn[ 

\title{Single-pulse stimulated Raman scattering spectroscopy}

\author{Hadas Frostig, Ori Katz, Adi Natan, and Yaron Silberberg}
\address{Department of Physics of Complex Systems, The Weizmann Institute of Science, Rehovot 76100, Israel}

\begin{abstract}
We demonstrate the acquisition of stimulated Raman scattering spectra with the use
of a single femtosecond pulse. High resolution vibrational spectra are obtained by
shifting the phase of a narrow band of frequencies in the broadband input pulse spectrum, using spectral shaping. The vibrational spectrum is resolved by examining the amplitude features formed in the spectrum after interaction with the sample. Using this technique, low frequency Raman lines ($<$100cm$^{-1}$) are resolved in a straightforward manner.
\end{abstract}

] 

\noindent Vibrational energy levels of molecules provide an
intrinsic contrast mechanism that can be exploited for spectroscopic
identification. Raman spectroscopy based on spontaneous Raman scattering is appealing due to its simplicity and readily interpretable spectra, but suffers from low
sensitivity~\cite{CARS_review}.
Coherent Raman scattering (CRS) techniques allow for
significantly higher sensitivity due to larger
cross-sections and forward scattering. As a result, CRS spectroscopy schemes
have been extensively studied over the past two decades. Until recently the dominant
CRS spectroscopy technique has been coherent anti-Stokes Raman
scattering (CARS). In the past several years, a different CRS process,
stimulated Raman scattering (SRS), has been proven suitable for high-sensitivity microscopy applications as well~\cite{Xie_science,Volkmer}.

In the common CARS and SRS schemes, a picosecond pump beam with frequency
$\omega_{p}$ and a picosecond Stokes beam with frequency
$\omega_{s}$ are focused onto the sample. When the beat frequency
$\omega_{p} - \omega_{s}$ matches a molecular vibrational frequency, $\omega_{vib}$,
third-order nonlinear polarizations are resonantly induced. The polarizations
act as sources of coherent fields at distinct frequencies that are
blue or red shifted by the vibration frequency from $\omega_{p}$ and $\omega_{s}$.
The result is the creation of two spectral sidebands around each input frequency,
at $\omega_{p}\pm\omega_{vib}$ and $\omega_{s}\pm\omega_{vib}$.
The vibrational spectrum is resolved by scanning the frequency of
$\omega_{p}$ or $\omega_{s}$ and observing the intensity of the Raman sidebands.
In practice, since CARS and SRS are third order nonlinear processes,
the generated field intensity is weak compared to the incoming
laser intensity. The straightforward approach, taken in CARS, is
thus to observe the generated electric field at a frequency different from
the input frequencies, $\omega_{cars}=2\omega_{p}-\omega_{s}$. This
allows the input frequencies to be filtered out, so that the weak
nonlinear signal can be detected. The CARS approach has proven
beneficial in spectroscopy, microscopy and biomedical imaging~\cite{CARS_review}.
A known limitation of CARS, however, is that it suffers from an intense
non-resonant background, which distorts the spectra and limits
detection sensitivity.

\begin{figure}[ht]
\centerline{\includegraphics[width=8.4cm]{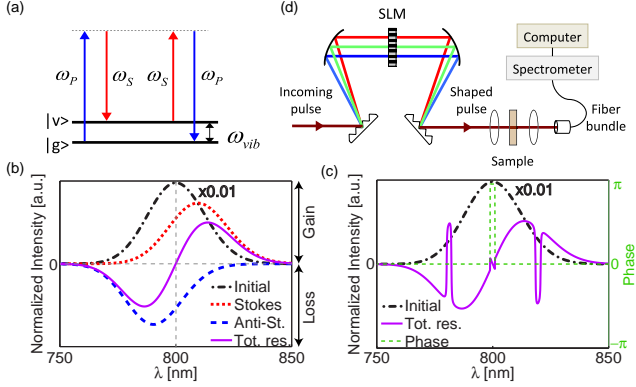}}
\caption{\label{SRS_setup_sim}{\footnotesize(a)~Energy level diagram of the stimulated Raman scattering process that creates signal at $\omega_{p}$.
(b)~The spectra of the input pulse (black), the Stokes (red) and anti-Stokes (blue) components of the resonant
signal (purple) generated from a sample with a single line at 300cm$^{-1}$. (c)~The spectrum (black) and phase (green) of a pulse shaped with
a $\pi$ phase gate pattern and the spectrum of the resonant signal (purple) it generates from the same sample. (d)~Schematic
of the experimental setup. The incoming pulses are dispersion compensated with a prism compressor (Femtolasres GmbH)
and then phase-shaped using a liquid-crystal spatial light modulator (Jenoptik Phase SLM-640). The beam is focused
onto the sample using a 15cm lens and the transmitted light is collected by a 0.5 NA lens and coupled into a CCD-based spectrometer (Jobin Yvon triax 320). For samples in powder form, the backscattered light is collected and coupled into the spectrometer.}}
\end{figure}

In contrast, the approach taken in SRS is to observe the generated electric field at a frequency that is
equal to one of the input frequencies, such as
$\omega_{p}=\omega_{s}+(\omega_{p}-\omega_{s})$. This process is
described in Fig.~\ref{SRS_setup_sim}(a). When doing so, one observes
the interference of the input field with the resonant and
non-resonant generated fields. The most significant terms of the
interference are given by:
\begin{eqnarray}\label{Interf}
\!\!\!\!I_{total}\!\!\!\!&=&\!\!\!\!|E_{in}+E_{nr}+E_{r}|^{2} \\
& \cong &\!\!\!\!I_{in} + 2|E_{in}||E_{nr}|\cos\phi_{nr} + 2|E_{in}||E_{r}|\cos\phi_{r} \nonumber
\end{eqnarray}
Where $E_{in}$ is the input electric field,
$E_{nr}$ and $E_{r}$ are the third order non-resonant and resonant electric field
responses respectively, and $\phi_{nr}$ and $\phi_{r}$ are their phases with respect to
$E_{in}$. The approximation in Eq. (\ref{Interf})
is valid in the perturbative limit commonly used in spectroscopy, where $|E_{in}| \gg |E_{r}|,|E_{nr}|$. The benefit of interference with the input field is
two-fold. First, as $E_{in}$ and $E_{nr}$
are in quadrature, the corresponding interference term vanishes ($\cos\phi_{nr} = 0$).
Second, $E_{in}$ and $E_{r}$, which are not in quadrature,
coherently interfere. Thus, the resonant signal detected in practice
corresponds to the third term in Eq. (\ref{Interf}). This results in heterodyne amplification of $E_{r}$ much like
in heterodyne CARS schemes\cite{Heterodyne_CARS}. The large background due to the input field
(first term in Eq. (\ref{Interf})) is avoided by modulating either the pump or Stokes
beams and measuring the output spectrum using lock-in techniques\cite{Xie_science}.
In accordance with theoretical analysis, previous SRS work has shown to provide
high-sensitivity and display negligible non-resonant background, enabling the
successful performance both in narrow-band\cite{Volkmer} and
multiplex schemes\cite{FSRS}. However, these schemes make use of
multi-beam configurations that require maintaining spatial and
temporal overlap of the beams and thus a complex experimental
setup. Development of a simpler SRS spectroscopy scheme can facilitate the
integration of SRS into practical biomedical use.

Here we propose an SRS scheme in which full spectral information can
be retrieved with the use of a single femtosecond pulse. In analogy
to single-pulse CARS schemes\cite{Phase_Contrast, Stand_off}, the use of a spectrally
broadband pulse provides both the pump and Stokes frequencies
necessary for excitation of a vibrational mode. Hence, the excitation is caused by all the
frequency pairs in the pulse with differences that match the
vibration frequency of the molecule. For the
case of a transform-limited pulse, the dynamics of
the system can be thought of as analogous to the multi-beam case,
where the excitation of the Raman mode imposes spectral sidebands on the pulse spectrum. The
resulting $E_{r}(\omega)$, for a pulse of an initial central frequency
$\omega_{0}$, has two components: a spectral replica of the incoming
pulse blue-shifted to a central frequency $\omega_{0}+\omega_{vib}$
(the anti-Stokes component) \cite{Fingerprint}
and another spectral replica red-shifted to a central frequency $\omega_{0}-\omega_{vib}$ (the Stokes
component). Due to the line shape of the molecular resonance, the Stokes and
anti-Stokes components are also phase-shifted with respect to the incoming field.
The phase-shift of the Stokes component results in positive $\cos\phi_{r}$ values corresponding to
stimulated Raman gain (red dotted line in Fig. \ref{SRS_setup_sim}(b)). The phase-shift of the anti-Stokes
component results in negative $\cos\phi_{r}$ values corresponding to stimulated Raman loss (blue dotted line
in Fig. \ref{SRS_setup_sim}(b)). For a broadband transform limited pulse,
the overall process results in a red-shift of the central frequency, making it hard to resolve
the vibrational spectrum (solid purple line in Fig. \ref{SRS_setup_sim}(b)).
One possible method of regaining chemical specificity is to instill a spectrally-narrow feature in
the pulse that will clearly mark the shifted position of the Stokes and anti-Stokes components.
Such a feature can be created by inducing a $\pi$ phase-shift in a narrow band of frequencies in the input pulse spectrum using spectral shaping as shown is Fig. \ref{SRS_setup_sim}(c) (this
phase pattern will be referred to as $\pi$ phase gate). The phase-shift
reverses the gain/loss properties of a narrow frequency band in the
shifted spectral replicas while hardly affecting the excitation of the
vibrational mode. This results in distinct, spectrally-narrow features in the measured output
spectrum that reveal the amount of frequency shift for each Raman line (solid purple line in Fig. \ref{SRS_setup_sim}(c)).
Alternatively, the effect of the $\pi$ phase gate can be considered in the time domain. The spectrally-narrow feature created by shaping in the frequency domain corresponds to a temporally-broad pulse in the time domain. The temporally-broad pulse probes the Raman level which was impulsively excited by the unshaped part of the pulse. Thus, the SPSRS scheme resembles the above mentioned multiplex SRS schemes.

To demonstrate single-pulse stimulated Raman scattering
spectroscopy, we measured the Raman spectra of several samples.
The experimental setup consisted of an amplified
Ti:sapphire laser emitting $\sim$30fs pulses centered at 795nm (corresponding to a
bandwidth of $\sim$50nm), a programmable
pulse-shaper based on a spatial-light modulator (SLM) and a
spectrometer (see Fig. \ref{SRS_setup_sim}(d)). The pulses were of
varying energies in the range 100nJ - 1$\mu$J at a 1KHz repetition rate. In order to eliminate
the constant input pulse background (first term in Eq. (\ref{Interf})),
which is several orders of magnitude larger than the resonant signal, we conducted a
differential measurement. The measurement was performed by comparing the spectra of
the light emerging from the sample for two slightly different
spectral locations of the $\pi$ phase gate. When subtracting the two,
not only does the constant input-pulse background vanish but also
the broad gain/loss pattern caused by the unshaped parts of the
pulse spectrum (see figures \ref{SRS_setup_sim}(b) and (c)). Therefore for frequencies
smaller than $\omega_{0}$, for example, where a transform-limited pulse would induce loss,
each Raman line is manifested as a narrow peak (loss converted to gain) and dip (subtracted peak) spectral feature on a rather flat background.

\begin{figure}[ht]
\centerline{\includegraphics[width=8.4cm]{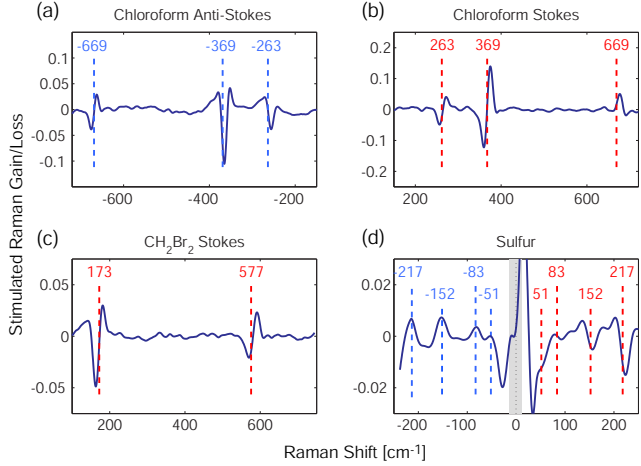}}
\caption{\label{Spectra} {\footnotesize SPSRS spectra of several samples. (a)~Chloroform anti-Stokes spectrum. (b)~Chloroform Stokes spectrum. (c)~Dibromomethane Stokes spectrum. (d)~Powdered sulfur spectrum. The phase gate area is marked in grey. Spectra (a)-(c) were resolved from the difference between two measurements with a 400ms integration time. Spectrum (d) was averaged over ten such difference measurements. The Raman vibrational lines are plotted on top for reference. The differences between the shapes of the Raman lines on the Stokes and anti-Stokes sides, as well as the spectral features in the vicinity of the gate in (d), are due to propagation effects.}}
\end{figure}

High-resolution Raman spectra of several materials are presented in
Fig. \ref{Spectra}. The measured spectra are in good agreement with the known
vibrational spectra of these materials. The Raman lines can be
easily identified either on the Stokes (Raman gain) or the anti-Stokes side
(Raman loss), allowing for flexibility in the experimental setup. We note that some of the resonant
features slightly deviate from the expected peak-dip shape.
This is the result of self-phase modulation acquired through propagation in the thick sample (1cm) we used. Nevertheless clear spectral features are visible enabling spectroscopic identification. The
demonstrated resolution is ~20cm$^{-1}$, limited by the resolution of
the SLM which dictates the minimal spectral width of the
phase gate that can be applied. Due to the simplicity of the measurement scheme, there is no need to filter out the excitation light. This is in contrast to many other filter-based Raman spectroscopy methods, in which the transition width of the filter is what limits the smallest Raman shift that can be measured. Consequently, low-frequency vibrational lines can be measured in a straightforward manner. As seen in Fig. \ref{Spectra}(d), several low lines of
Sulfur at 217, 152 and 83 cm$^{-1}$ are easily discerned and a feature indicating the 51cm$^{-1}$ is noticeable on the Stokes side, all in agreement with previous work \cite{Sulfur_raman}.
Generally, the lowest line that can be detected using this scheme is limited by the spectral width of the $\pi$ phase gate. However, in thick samples, propagation-related effects can
cause spectral changes in the vicinity of the $\pi$ phase gate \cite{Probe_enhancement} and impose a
higher bound on the lowest detectable line, as in the Fig. \ref{Spectra}(d).

\begin{figure}[ht]
\centerline{\includegraphics[width=8.4cm]{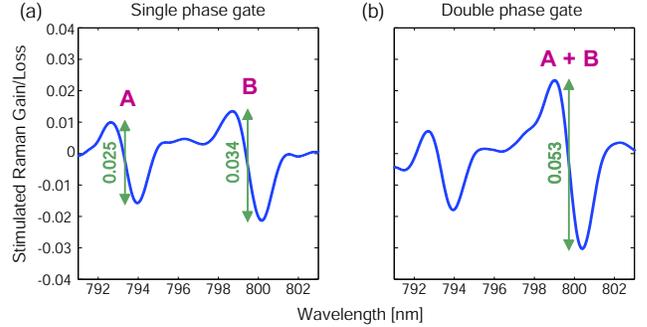}}
\caption{\label{Coh_add} {\footnotesize Coherent addition of Raman lines in Carbon Tetrachloride. (a)~The
spectrum generated by a pulse shaped with a single phase gate at 780nm, showing
two Raman lines: 218cm$^{-1}$ (at 793nm) and 314cm$^{-1}$ (at 799nm). (b)~The spectrum generated by a pulse
shaped with two phase gates spaced by the difference between the two vibrational frequencies, 96cm$^{-1}$ (gates at 780nm and 786nm). The 314cm$^{-1}$ line of the first phase gate and the 218cm$^{-1}$
line of the second phase gate both occur at 799nm. The resulting peak-to-dip difference approximately equals the
sum of the peak-to-dip differences of the two Raman lines in (a).}}
\end{figure}

Another advantageous feature of SPSRS over conventional picosecond SRS
arises when the desired goal is detection of a predetermined
substance. When applying several $\pi$ phase
gates to the spectrum spaced by the differences between the vibrational frequencies of
the molecule, the peak-dip features from all the gates are generated at
the same frequency. Due to the coherent nature of the process, the signals
coherently combine to create a single larger spectral feature. This feature
indicates the level of correlation between the measured spectrum and
the known spectrum of a substance \cite{All_optical}. The benefits are enhanced
signal-to-noise ratio (SNR) compared to each single line, as well as elimination of the need for post-processing of the spectrum. The ability to coherently add lines using SPSRS is demonstrated
in the spectrum of carbon tetrachloride, as shown in Fig. \ref{Coh_add}. The single-gate spectrum Fig. \ref{Coh_add}(a) reveals two resonant lines which create features at 793nm and 799nm. In the double-gated spectrum shown in Fig. \ref{Coh_add}(b) the two features are combined at 799nm.
The combined feature has a peak-to-dip difference that nearly equals (90\%) the sum of the two
individual lines, representing an appreciable enhancement of the SNR.
The feature can be further enlarged by optimizing the relative phase of the gates and
their spectral locations either manually or through the use of an adaptive algorithm \cite{All_optical}.

In conclusion, we have demonstrated vibrational spectra acquisition
using a single-pulse SRS scheme. Spectral shaping enables full control
over the interference of the input field and the generated resonant electric field,
facilitating the creation of narrow spectral features
that indicate the frequencies of the Raman lines. Using this
method, all vibrational levels within the detection range can be
simultaneously identified in a similar fashion to multiplex SRS
schemes but with a single beam setup. Two unique features of the SPSRS
scheme have been demonstrated, the ability to distinguish low-lying
Raman lines and the ability to coherently add the signal from several
lines for an improved spectroscopic fingerprint. Spectroscopy of low-lying Raman lines is
a useful tool in various research fields. Examples include monitoring electrically distinct carbon
nanotubes \cite{nanotubes} as well as studying the hydration dynamics of
DNA films \cite{DNA}. Furthermore, by employing fast shaping techniques together with lock-in detection \cite{LC2D} the sensitivity of our setup can be substantially increased and SPSRS can become an attractive scheme for microcopy and biological imaging.

The authors thank MOST, NATO and the Crown Photonic Center for their financial support.

\end{document}